\documentclass[11pt]{report} 
\usepackage{graphics}
\input{epsf}
\begin{document}
\title{Introduction To Monte Carlo Algorithms 
}
\author{Werner Krauth\thanks{werner.krauth@ens.fr, 
http://www.lps.ens.fr/$\tilde{\;}$krauth}
\footnote{
Note added (2006): Material discussed in these lecture notes is treated in
similar style but with a broader perspective in my book
"Statistical Mechanics: Algorithms and Computations"
(Oxford University Press, 2006).
For excerpts and further information, see the book's website: 
http://www.phys.ens.fr/doc/SMAC.}\\
CNRS-Laboratoire de Physique Statistique\\
Ecole Normale Sup\'{e}rieure\\
24, rue Lhomond\\
F-75231 Paris Cedex 05, France\\
\\ 
\\ 
(``Advances in Computer Simulation'',
Lecture Notes in Physics \\ 
J. Kertesz and I. Kondor, eds (Springer Verlag, 1998))
\\ 
\\ 
}
\date{December 20, 1996} 
\maketitle
\begin{abstract}
In these lectures, given in '96 summer schools  in Beg-Rohu (France)
and Budapest, I discuss the fundamental principles of thermodynamic
and dynamic Monte Carlo methods in a simple light-weight fashion.
The keywords are \textsc{Markov chains, Sampling, Detailed Balance, A Priori
Probabilities, Rejections, Ergodicity, ``Faster than the clock algorithms"}.

The emphasis is on \textsc{Orientation}, which is difficult to obtain
(all the mathematics being simple). A firm sense of orientation helps to avoid
getting lost, especially if you want to leave safe trodden-out paths
established by common usage.

Even though I will remain quite basic (and, I hope, readable), 
I make  every effort to drive home the essential messages, which are
easily explained: the crystal-clearness of detail balance, the main
problem with Markov chains, the great algorithmic freedom, both in 
thermodynamic and dynamic Monte Carlo, and the fundamental differences
between the two problems.
\end{abstract}

\chapter{Equilibrium Monte Carlo methods}

\section{A Game in Monaco}
\label{Monaco}
\begin{figure}[htbp] \unitlength 1cm
\begin{center}
\begin{picture}(6.8,5.)
\put(0.0,0){\epsfbox{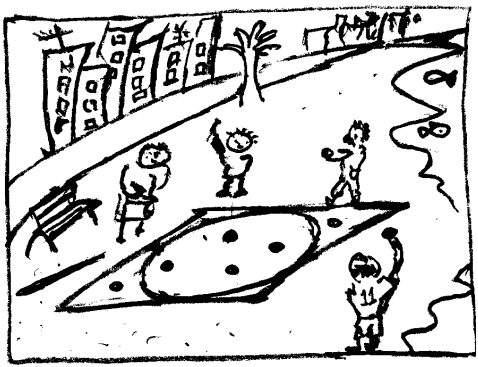}}
\end{picture}
\end{center}
\caption
{Children at play on the beaches of Monaco. They spend their afternoons
calculating $\pi$ by a method which can be easily extended to general 
integrals.}
\label{1children}
\end{figure}

The word ``Monte Carlo method'' can be traced back to a game very popular
in Monaco. It's not what you think, it's mostly a children's pass-time
played on the beaches. On Wednesdays (when there is no school) and  
on weekends, they get together, pick up a big stick, draw a circle and 
a square as shown in figure~\ref{1children}. 
They fill their pockets with pebbles 
\footnote{pebble is 
{\em calculus} in Latin.}. Then they stand around, 
close their eyes, and throw the pebbles randomly in the direction of the
square. Someone keeps track of the number of pebbles which hit the 
square, and which fall within the circle (see figure~\ref{1children}). 
You will easily 
verify that the ratio of pebbles in the circle to the ones in the
whole square should come out to be $\pi/4$, so there is much excitement
when the $40$th, $400$th, $4000$th is going to be cast.

This breath-taking game is the only method I know to compute the number
$\pi$ to 
arbitrary precision without using fancy measuring devices (meter, balance)
or advanced mathematics (division, multiplication, trigonometry).
Children in Monaco can pass the whole afternoon at this game. You are 
invited \footnote{cf. ref.~\cite{Press} for an unsurpassed discussion of random 
numbers, including all practical aspects.}
to write a little program to simulate the game. If you have
never written a Monte Carlo program before, this will be your first one.
You may also recover ``\texttt{simplepi.f}'' from my WWW-site.

Lately, big people in Monaco have been getting into a similar game.
Late in the evenings, when all the helicopters are safely stowed away, 
\begin{figure}[htbp] \unitlength 1cm
\begin{center}
\begin{picture}(8.,5.)
\put(0.0,0){\epsfbox{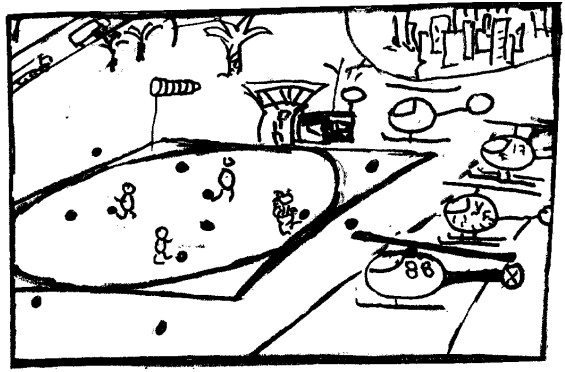}}
\end{picture}
\end{center}
\caption
{Adults at play on the Monte Carlo heliport. Their method to calculate $\pi$ 
has been extended to the investigation of liquids, suspensions, and lattice
gauge theory.}
\label{1grownups}
\end{figure}
they get together on the local heliport (cf. figure~\ref{1grownups}), 
which offers the same basic layout
as in the children's game. They fill their expensive Herm\`{e}s handbags  
with pebbles, but since the field is so large, they play a slightly different
game: they start somewhere on the field, close their eyes, and then 
throw the little stone  in a random direction. Then they walk to where 
the first stone landed, 
take a new one out of their handbag, and start 
again. You will realize that using this method, one can also sweep out 
evenly the heliport square, and compute
the number $\pi$. You are invited to write a $10$-line program 
to simulate the heliport game - but it's not completely trivial.

Are you starting to think that the discussion is too simple? If that is
so, please consider the lady at c). She just flung a pebble to c'),
which is outside the square. What should she do?
\begin{enumerate}
\item simply move on at c)
\item climb over the fence, and continue until, by accident, she will 
   reintegrate the heliport
\item other: \framebox[10cm]{\rule{0cm}{.5cm}}
\end{enumerate}

A correct answer to this question will be given on page~\pageref{answerdb}, 
it contains the 
essence of the concept of detailed balance. Many Monte Carlo programs
are wrong because the wrong alternative was programmed.   

The two cases - the children's game and the grown-ups' game are perfect 
illustrations of what is going on in Monte Carlo\ldots and in Monte Carlo
algorithms. In each case, one is interested in evaluating an integral 
\begin{equation}
\int_{x,y\;\epsilon \framebox[.2cm]{} } dx dy \pi(x,y) f(x,y)
\end{equation}
with a {\em probability density} $\pi$ which, in our case is the square 
\begin{equation}
\pi(x,y)=  \left\{ \begin{array}{ll}
       1 & \mbox{if $|x| <1$ and $|y| <1$}\\
       0 & \mbox{otherwise}
                 \end{array}
        \right. 
\label{density}
\end{equation}
and a function $f$ (the circle)
\begin{equation}
f(x,y)=  \left\{ \begin{array}{ll}
       1 & \mbox{if $x^2 + y^2 < 1$}\\
       0 & \mbox{otherwise}
                 \end{array}
        \right. 
\label{function}
\end{equation}
Both the children and the grown-ups fill the square with a constant
density of pebbles (corresponding to $\pi(x,y)=1.)$, one says that
they {\em sample} the function $\pi(x,y)$ on the basic square.
If you think about it you will realize that the number $\pi$ can be 
computed in the two games only because the area of the basic square is
known. If this is not so, one is reduced to computing the ratio of the 
areas of the circle and of the square, i.e., in the general case, the 
ratio of the integrals: 
\begin{equation}
\int_{x,y\;\epsilon \framebox[.2cm]{} } dx dy \pi(x,y) f(x,y) /
\int_{x,y\;\epsilon \framebox[.2cm]{} } dx dy \pi(x,y)  
\label{ratio_of_integs}
\end{equation}

Two basic approaches are used in the Monte Carlo method:
\begin{enumerate}
\item direct sampling (children on the beach)
\item Markov-chain sampling (adults on the heliport)
\end{enumerate}

Direct sampling is usually like \textsc{Pure Gold}, it means that you 
are able to 
call a subroutine which provides an independent hit at your distribution 
function $\pi(x,y)$. This is exactly what the kids do 
whenever they get a new pebble
out of their pockets. 

\begin{figure}[htbp] \unitlength 1cm
\begin{center}
\begin{picture}(6.,4.)
\put(0.0,0){\epsfbox{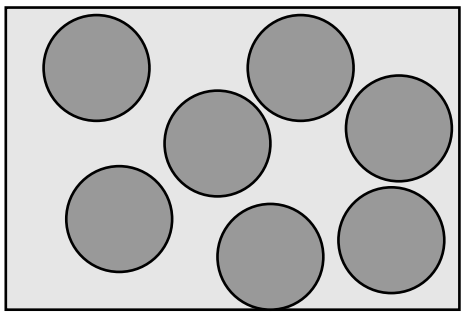}}
\end{picture}
\end{center}
\caption
{The Coin in a shoe-box problem (hard disks) has occupied 
Monte Carlo workers since 1953. There is no direct-sampling algorithm, 
and Markov-chain sampling is extremely tedious at some densities.
}
\label{1coin}
\end{figure}

Most often, there is no way to do direct sampling in a reasonable
manner. One then resorts to Markov-chain sampling, as the adults
on the heliport. Almost all physical systems are of this class.
A famous example, which does not oblige us to speak of energies, 
hamiltonians etc. has occupied more than a generation
of physicists. It can easily be created with a shoe-box, and a number of 
coins (cf figure~\ref{1coin}): how do you generate (directly sample) 
random configurations of the coins such that they don't overlap? Imagine the 
$2 N$-dimensional configuration
space of $N$ non-overlapping coins in a shoe-box.  Nobody has found a
subroutine which would directly sample this configuration space, i.e.
create any of its members with equal probability.
First-time listeners often spontaneously propose a method called 
\textsc{Random Sequential Adsorption}: deposit a first coin at a random
position, then a second, etc, (if they overlap, simply try again). 
Random Sequential Adsorption will be dealt with in detail, but in 
a completely different context, on page~\pageref{depos}. Try to 
understand that this has nothing to do with finding a random non-overlapping
configuration of the coins in the shoe-box (in particular, the average maximum
density of random sequential deposition is much smaller than the close
packing density of the coins).

Direct sampling is usually impossible - and that is the basic frustration
of Monte Carlo methods (in statistical physics). In contrast,  the 
grown-ups' game can be played in any situation (\ldots already on the heliport 
which is too large for direct sampling). Markov chain  sampling has been 
used in an uncountable number of statistical physics models, in the 
aforementioned coin-in-a-box problem, the Hubbard model, etc etc. Usually
it is a very poor, and extremely costly substitute to what we 
really want to do.

In the introduction, I have talked about Orientation, and to get oriented
you should realize the following:
\begin{itemize}
\item In the case of the children's game, you need only a few dozen 
pebbles (samples) to get a physicist's idea of the value of $\pi$, which 
will be sufficient for most applications. Exactly the same is true
for some of the most difficult statistical physics problems. A few 
dozen (direct) samples of large coin-in-the-box problems at any density 
would be sufficient to resolve long-standing controversies. (For a random
example, look up ref. \cite{Lee}, 
where the analysis of a monumental simulation relies on an effective 
number of $4-5$ samples).

Likewise, a few 
dozen direct samples of the Hubbard model of strong fermionic correlation
would give us important insights into superconductivity. 
Yet, some of the largest computers in the world are churning away day after
day on 
problems as the ones mentioned. They all try to bridge the gap between 
the billions of Markov chain samples
and the equivalent of a few random flings in the children's game.

\item It is only after the problem to generate independent samples
by Markov chains {\em at all} is understood, that we may start to 
worry about the slow convergence of mean-values. This is already 
apparent in the children's game - as in every  measurement in 
experimental physics - : the precision of the numerical value decreases
only as $1/\sqrt{N}$, where $N$ is the number of independent measurements.
Again, let me argue against the ``mainstream'' that
the absolute precision 
discussion is not nearly as important as it may seem: you are not
always interested in computing your specific heat to five significant
digits before putting your soul to rest. 
In daily  practice of Monte Carlo it is usually more critical to be absolutely
sure that the program has given some independent samples 
\footnote{{\em i. e.} that
it has decorrelated from the initial configuration}
rather than that there are millions and billions of them.

\item It is essential to understand that a long Markov chain, even 
if it produces only a small number of independent samples (and thus a 
very approximative result) is usually extremely sensitive to bugs,
and even small irregularities of the program. 
\end{itemize}

\section{The Toddlers' algorithm and detailed balance}
\label{toddler}
\begin{figure}[htbp] \unitlength 1cm
\begin{center}
\begin{picture}(3.,3.)
\put(0.0,0){\epsfbox{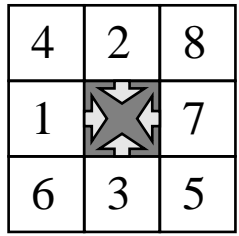}}
\end{picture}
\end{center}
\caption
{This well-known puzzle is the starting point for our theoretical analysis
of the Monte Carlo method.}
\label{1puzzle}
\end{figure}

\begin{figure}[htbp] \unitlength 1cm
\begin{center}
\begin{picture}(6.8,5.)
\put(0.0,0){\epsfbox{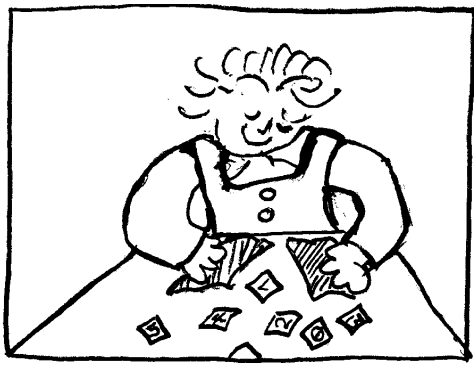}}
\end{picture}
\end{center}
\caption
{This is a direct sampling algorithm for the puzzle game. }
\label{1toddler}
\end{figure}

We have yet to determine what the lady at point
c) in the heliport game should do. It's a difficult question, and full
of consequences,  and we don't want to give her any wrong advice.
So, let's  think, and  analyze first a similar, discretized game, 
the well-known puzzle shown in figure~\ref{1puzzle}.  Understanding this
puzzle will allow us to make a definite recommendation.
The task is now to create a perfect scrambling  algorithm which generates 
any possible configuration of the puzzle with equal probability.
One such method, the toddlers' algorithm, is
well known, and illustrated in figure~\ref{1toddler} together with one of 
its  inventors. 

The theoretical baggage picked up in the last few sections allows
us to class this method without hesitation among the 
direct sampling methods (as the children's game before), since it creates 
an independent instance each time the puzzle is broken apart. 

We are rather interested in a grown-up people's algorithm, which respects the 
architecture of the game \footnote{Cf. the discussion on 
page~\pageref{ergodbreak} for an important and 
subtle difference between the toddlers' algorithm and any incremental
method.}.   

What would \textsc{You} do?  Almost certainly, you would hold the puzzle
in your hands and - at any time step - move the empty square in a 
random direction. The detailed balance condition which you will 
find out about in this section shows that this completely natural
sounding algorithm is \textsc{wrong}.

If the blank square is in the interior (as in figure~\ref{1puzzle})
then the algorithm should clearly be to choose one of the four
possible directions 
\makebox[.6cm]{\rule{-.3cm}{.3cm}$\uparrow$}
\makebox[.6cm]{\rule{-.3cm}{.3cm}$\rightarrow$}
\makebox[.6cm]{\rule{-.3cm}{.3cm}$\downarrow$}
\makebox[.6cm]{\rule{-.3cm}{.3cm}$\leftarrow$}
with equal probability, and to move the empty square.

As in the heliport game, we have to analyze what happens at the boundary
of the square
\footnote{periodic boundary conditions are \textsc{not} allowed.}.
\begin{figure}[htbp] \unitlength 1cm
\begin{center}
\begin{picture}(6.9,4.3)
\put(0.0,0){\epsfbox{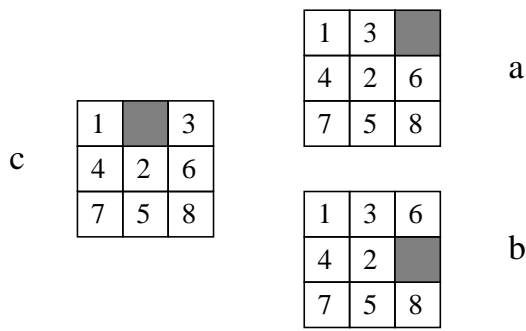}}
\end{picture}
\end{center}
\caption
{The corner configuration $a$ is in contact with the configurations 
$b$ and $c$}
\label{1detailed}
\end{figure}
Consider the corner configuration $a$, which communicates
with the configurations $b$ and $c$, as shown in figure~\ref{1detailed}. 
If our algorithm
(yet to be found) generates the configurations $a$, $b$ and $c$ with 
probabilities $\pi(a)$, $\pi(b)$, and $\pi(c)$, respectively
(we impose them to be the same), we can derive a simple rate equation
relating the $\pi$'s to the transition probabilities $p(a\rightarrow b)$
etc. Since $a$ can only be generated from $b$, $c$, or from itself, we have
\begin{equation}
\pi(a)= \pi(b) p(b\rightarrow a) +  
\pi(c) p(c\rightarrow a) + \pi(a) p(a\rightarrow a)
\end{equation}
this gives
\begin{equation}
\pi(a)[1- p(a\rightarrow a)]  =  
\pi(b) p(b\rightarrow a) + \pi(c) p(c\rightarrow a)
\label{at_a}
\end{equation}
We write the condition which tells us that the  empty square, once at $a$,
can stay at $a$ or move to $b$ or $c$:
\begin{equation}
1 = p(a\rightarrow a) + p(a\rightarrow b) + p(a\rightarrow c)
\label{sumofprob}
\end{equation}
which gives
$[1 - p(a\rightarrow a)] = p(a\rightarrow b) + p(a\rightarrow c)$.
This formula, introduced in eq. (\ref{at_a}) yields
\begin{equation}
 \pi(a) \underbrace{p(a\rightarrow b) +  \pi(a) \overbrace{p(a\rightarrow c) 
  =   \pi(c)} p(c\rightarrow a) + \pi(b)} p(b\rightarrow a)
\end{equation}
We can certainly satisfy this equation if we equate the braced terms
separately:
\begin{equation}
\pi(a) p(a\rightarrow b) = \pi(b) p(b\rightarrow a) \;\;\mbox{etc.}
\label{detail}
\end{equation}
This equation is the celebrated \textsc{Condition of detailed Balance}.

We admire it for awhile, and then pass on to its first application, in 
our puzzle. There, we impose equal probabilities for all accessible 
configurations, i. e.  $p(a\rightarrow b) = p(b\rightarrow a)$, etc, 
and the only simple way to connect the proposed algorithm for the 
interior of the square with the boundary is  to \textsc{postulate} an 
equal probability of $1/4$ for any possible move. Looking at 
eq.~(\ref{sumofprob}),
we see that we have to allow a probability of $1/2$ for    
$p( a \rightarrow a)$.  The consequence of this analysis is
that  to maximally scramble the puzzle, we have to be immobile with
a probability $1/2$ in the corners, and with probability $1/4$ on the
edges. 

We can assemble all the different rules in the following algorithm:
At each time step $t=0,1,\ldots$
\begin{enumerate}
\item choose one of the four directions 
\makebox[.6cm]{\rule{-.3cm}{.3cm}$\uparrow$}
\makebox[.6cm]{\rule{-.3cm}{.3cm}$\rightarrow$}
\makebox[.6cm]{\rule{-.3cm}{.3cm}$\downarrow$}
\makebox[.6cm]{\rule{-.3cm}{.3cm}$\leftarrow$}
with equal probability.
\item move the blank square into the corresponding direction {\em if
possible}. Otherwise, stay where you are, but advance the clock.
\end{enumerate}
I have presented this puzzle algorithm in the hope that you will 
\textsc{not} believe that it is true. This may take you to 
scrutinize the detailed balance condition, and to understand it better.

\begin{figure}[htbp] \unitlength 1cm
\begin{center}
\begin{picture}(7.,5.)
\put(0.0,0){\epsfbox{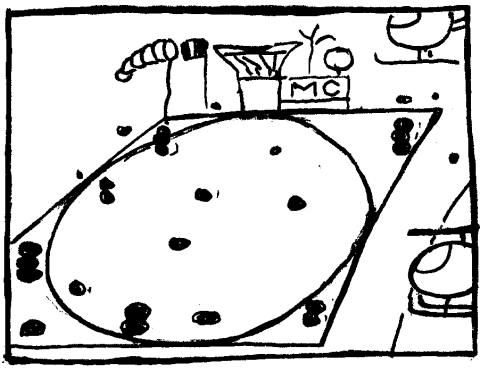}}
\end{picture}
\end{center}
\caption
{Landing pad of the Monte Carlo heliport after all the players have gone
home. Notice the piles at the 
edges, particularly close to the corners.}
\label{1afterthegame}
\end{figure}

Recall that the puzzle game was in some respect a discretized version 
of the heliport game, and you will now be able to answer by analogy the 
question of the fashionable lady \label{answerdb} at c).
Notice that there is already a pebble at
this point. The lady should now pick two more stones out of  her handbag, 
place  one of them
on top of the stone already on the ground, and use the remaining one to
try a new fling. If this is again an
out-fielder, she should pile one more stone up etc. If you look onto
the heliport after the game has been going on for awhile, you will 
notice a strange pattern of pebbles on the ground 
(cf. figure~\ref{1afterthegame}). 
There are mostly single stones in the interior, and some more or
less high piles as you approach the boundary, especially the corner. 
Yet, this is the most
standard way to enforce a constant density  $\pi(x,y) = const$. 

It has happened that 
first-time listeners of these sentences are struck with 
utter incredulity. They find the piling-up of stones absurd
and conclude that {\em I} must have gotten the story 
wrong. The only way to reinstall confidence is to show 
simulations (of ladies flinging stones) which do and do not 
pile up stones on occasions. You are invited to do the same 
(start with one dimension). 

In fact, we have just seen the first instance of a \textsc{Rejection}, 
which, as announced, is a keyword of this course, and of the Monte
Carlo method in general. The concept of a rejection is so fundamental
that it is worth discussing it in the completely barren context 
of a Monaco airfield. Rejections are the basic method by which 
the Monte Carlo enforces a correct density $\pi(x,y)$ on the square, 
with an algorithm (random ``flings") which is not particularly suited
to the geometry. Rejections are also wasteful (of pebbles), and expensive
(you throw stones but in the end you just stay where you were). 
We will deepen our understanding of rejections considerably in the 
next sections.
 
We have introduced the rejections by a direct inspection of the 
detailed balance condition. This trick  has been elevated to 
the status of a general method in the socalled Metropolis algorithm 
\cite{Metrop}. 
There, it is realized that the detailed balance condition 
eq.~(\ref{detail}) is verified if one uses
\begin{equation}
     P(a \rightarrow b) = \min(1, \frac{\pi(b)}{\pi(a)})
\label{Metrop}
\end{equation}
What does this equation mean? Let's find out in the case of the 
heliport: Standing at a) (which is inside the square, i.e. $\pi(a)=1$), 
you throw your pebble in a random 
direction (to $b$). Two things can happen: Either $b$ is inside
the square ($\pi(b)=1$), and eq.~(\ref{Metrop}) tells you to accept 
the move with
probability $1$, or $b$ is outside ($\pi(b)=0$), and you should
accept with probability $0$, i.e. reject the move, and stay where you are.

After having obtained an operational understanding of the 
Metropolis algorithm,
you may want to see whether it is \textsc{correct} in the general case.
For once, there is rescue through bureaucracy, for the  
theorem can be checked by a bureaucratic procedure: simply fill
out the following form:\\ 
\begin{center}
\begin{tabular}{||r|c|c|c||} \hline
case                     & $\pi(a) > \pi(b)$ &  $\pi(b) > \pi(a)$ & \\ \hline
$P(a\rightarrow b)$      &                   &                    &$1$\\
                                                                  \cline{2-3}
$\pi(a)P(a\rightarrow b)$&                   &                    &$2$\\ \hline
                 \hline
$P(b\rightarrow a)$      &                   &                    &$3$\\ 
                                                                  \cline{2-3}
$\pi(b)P(b\rightarrow a)$&                   &                    &$4$\\ \hline
\end{tabular}
\end{center}
Fill out, and you will realize for both columns that the second and 
forth rows are identical, as stipulated by the detailed balance condition.
You have understood all there is to the Metropolis algorithm.

\section{To Cry or to \textsc{Cray}}
Like a diamond tool, the Metropolis algorithm allows you to 
``grind down'' an arbitrary density of trial movements
(as the random stone-flings) into the chosen stationary probability density
$\pi(x,y)$. 

To change the setting we discuss here how a general  classical
statistical physics model with an arbitrary high-dimensional energy  
$E(x_1,
x_2,\ldots,x_N$) is simulated (cf the classic reference \cite{Binder}). 
In this case, the probability density is the 
Boltzmann measure
$\pi(x_1,...,x_N)= \exp(-\beta E)$, and the physical expectation values
(energies, expectation values) are given by formulas as in 
eq.~(\ref{ratio_of_integs}).
All you have to do
(if the problem is not too difficult) is to \ldots
\begin{itemize}
\item set up a (rather arbitrary) displacement rule, which should 
generalize the random stone-flings of the heliport game. For example, 
you may from an initial configuration $x_1,x_2,\ldots,x_N$ go to a new 
configuration by choosing an arbitrary dimension i, and doing the random 
displacement on $x_i \rightarrow x_i+ \delta x$, with $\delta x$ between
$-\epsilon$ and $ + \epsilon$. Discrete variables are treated just as
simply.
\item Having recorded the energies $E^a$ and $E^b$ at the 
initial point $a$ and the 
final point $b$, you may use the Metropolis algorithm to compute the 
probability, $p$,  to \textsc{actually} go there. 
\begin{equation}
p(a\rightarrow b) = \min[1,\exp( -\beta(E^b - E^a)])
\end{equation}
This is implemented
using a single uniformly distributed random number $0 < ran<1$, and 
we move our system to $b$ under the condition that 
$ran < p(a \rightarrow b)$, as shown in the figure.
\begin{figure}[htbp] \unitlength 1cm
\begin{center}
\begin{picture}(4.,2.9)
\put(.0,0.0){\epsfbox{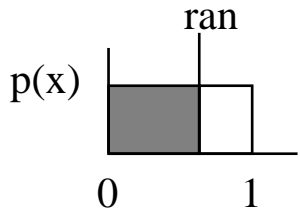}}
\end{picture}
\end{center}
\caption
{The probability for a uniformly distributed random number $0<ran<1$ to 
be smaller than $p(a \rightarrow b)$ is $\min[1,p(a \rightarrow b)]$.
A typical Monte Carlo program contains one such comparison per step.}
\label{1mcdecision}
\end{figure}
\item You notice that (for continuous systems) the remaining liberty in 
your algorithm is to 
adjust the value of $\epsilon$. The time-honored procedure consists in 
choosing $\epsilon$ such that about half of the moves are rejected. This is of
course just a golden rule of thumb. The ``$<p(\epsilon)>\sim 1/2$'' rule, as
it may be called, \textsc{does} in general insure quite a  
quick diffusion of your particle across configuration space. If you want to 
do better, you have to monitor the speed of your Monte Carlo diffusion, 
but it is usually not worth the effort.
\end{itemize}

As presented, the Metropolis algorithm is extremely powerful, and many 
problems are adequately treated with this method. The method is so 
powerful that for many people the theory of Monte Carlo stops right 
after equation (\ref{detail}).
All the rest is implementation details, data collection, and the 
adjustment of $\epsilon$, which was just discussed. 

Many problems, especially in high dimensions, nevertheless defy this rule. 
For these problems, the programs written 
along the lines of the one presented above will run properly, but
have a very hard time generating independent samples at all. These 
are the problems on which one is forced to either give up or compromise:
use smaller sizes than one wants, take risks with the error analysis etc.
The published papers, over time, present much controversy, shifting results,
and much frustration on the side of the student executing the calculation.

Prominent examples of difficult problems are phase transitions 
in general statistical 
models, Hubbard model, Bosons, and disordered systems. The 
strong sense of frustration can best be retraced in the case of the 
hard sphere liquid, which was first treated in the original $1953$ publication 
introducing the Metropolis algorithm, and which has since generated
an unabating series of unconverged calculations, and heated controversies.

\begin{figure}[htbp] \unitlength 1cm
\begin{center}
\begin{picture}(10.6,4.1)
\put(0.0,0){\epsfbox{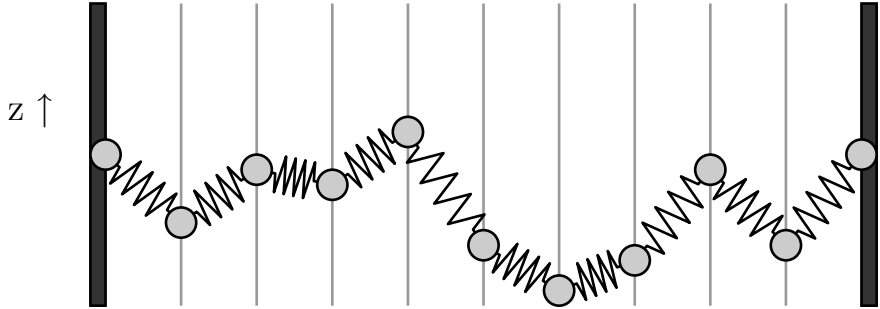}}
\put(-1.,2.5){\mbox{{\Large z$\;\uparrow$}}}
\end{picture}
\end{center}
\caption
{Chain of coupled springs which serves as an example of a large-dimensional
Monte Carlo problem. }
\label{1oscillators}
\end{figure}

A very useful system to illustrate a difficult simulation, shown in 
figure~\ref{1oscillators}
is the chain of $N$ springs with an energy 
\begin{equation}
E = \frac{(z_1-z_0)^2}{2} + \frac{(z_2-z_1)^2}{2} + \ldots +
\frac{(z_{N+1}-z_N)^2}{2}
\label{oscillators}
\end{equation}
$z_0$ and $z_{N+1}$ are fixed at zero, and $z_1,\ldots z_N$ are the 
lateral positions of the points, the variables of the system.
How do we generate distributions according to $\exp(-\beta E)$? 
A simple algorithm
is very easily written 
(you may recover ``\texttt{spring.f}'' from my WWW-site). 
According
to the recipe given above you simply choose a random bead and displace
it by a small amount. The whole program can be written in about $10$ lines.
If you experiment with the program, you can also try to optimize the value 
of $\epsilon$ (which will come out such that the average change
of energy in a single move is  $\beta | E^b - E^a| \sim 1$).

What you will find out is that the program works, but is extremely slow.
It is so slow that you want to cry or to \textsc{Cray} (give up or 
use a supercomputer), and both would be ridiculous.

Let us analyze why a simple  program for the coupled springs is so 
slow. The reason are the following
\begin{itemize}
\item Your ``$<p(\epsilon)> \sim 1/2$'' rule fixes the step size, which is 
necessarily very small. 
\item The distribution of, say,  $z_{N/2}$, the middle bead, as
expressed \textsc{in units of $\epsilon$} 
is very large, since the whole chain will have a lateral extension of 
the order of $\sqrt{N}$. It is absolutely essential to realize that the
distribution can never be intrinsically wide, but only in units of the 
imposed step size (which is a property of the algorithm).
\item It is very difficult to sample a large distribution
with a small stepsize, in the same way as it takes a very large bag
of stones 
to get an approximate idea of the numerical value of $\pi$ if the 
heliport is much larger than your throwing range (cf. figure~\ref{1slow}).
\begin{figure}[htbp] \unitlength 1cm
\begin{center}
\begin{picture}(10.,5.1)
\put(0.0,0.3){\epsfbox{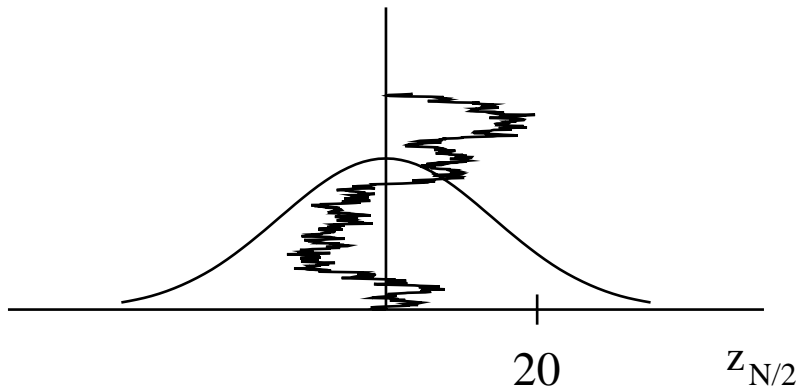}}
\end{picture}
\end{center}
\caption
{ The elastic string problem with $N=40$ is quite a difficult
problem because the distribution of, e g, $z_{N/2}$ is large.
The simulation path shown in the figure corresponds to $400.000$ steps.
 }
\label{1slow}
\end{figure}
\end{itemize}
The difficulty to sample a wide distribution is  the  basic 
reason why simulations can have difficulties converging.

At this point, I often encounter the remark: why can't I move several
interconnected beads independently, thus realizing a large move? 
This strategy is useless.\label{simultaneous}
In the big people's game, it corresponds to trying to save every second
stone by throwing it into a random direction, fetching it (possibly 
climbing over the fence), 
picking
it up and throwing it again. You don't gain anything since you already 
optimized the throwing range before. You already 
had an important probability of rejection, which will now become 
prohibitive. The increase in the rejection probability will more than
annihilate the gain in stride.

As set up, the thermal spring problem is difficult because the many 
degrees of freedom $x_1,\ldots x_N$ are strongly coupled. Random
$\epsilon$-moves are an extremely poor way to deal with such systems. 
Without an improved strategy for the attempted  moves, the program very quickly 
fails to converge, {\em i e} to produce even a handful of independent 
samples. 

In the coupled spring problem, there are essentially two ways
to improve the situation. The first consists in using a basis transformation,
in this case in simply going to 
Fourier space. This evidently decouples the degrees of freedom. You 
may identify the Fourier modes  which have a chance of being excited.
If you write a little program, you will very quickly master a popular
concept called ``Fourier acceleration''. An exercise of unsurpassed value
is to extend the program to an energy which contains a small additional 
anisotropic coupling of the springs and treat it with both algorithms.
Fermi, Pasta and Ulam, in $1945$, were the first to simulate the anisotropic 
coupled chain problem on a computer (with a deterministic algorithm) 
and to discover extremely slow thermalization. 

The basis transformation is a specific method to allow large moves.
Usually, it is however impossible to find such a transformation.
A more general method to deal with difficult interacting problems consists
in  isolating subproblems which you can more or less  sample easily
and solve exactly.
The \textsl{a priori} information gleaned from this analytical work can then
be used to propose exactly those (large) moves which are compatible with
the system. The proposed moves are then rendered correct by means of a
generalized Monte Carlo  algorithm. A modified Metropolis 
rejection remains, it corrects the ``engineered'' density into the
true stationary probability. We will first motivate this very important
method in the case of the coupled chain example, and then give the 
general theory, and present a very important application to spin models.

\begin{figure}[htbp] \unitlength 1cm
\begin{center}
\begin{picture}(6.7,5.5)
\put(0.0,0){\epsfbox{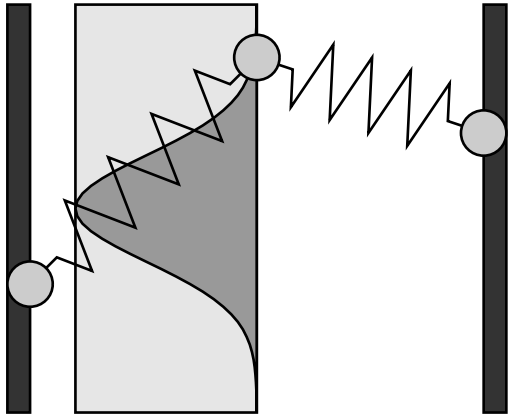}}
\end{picture}
\end{center}
\caption
{Here we analyze part of the coupled spring simulation, with $z_{i-1}$
and $z_{i+1}$ immobilized. The large rejection rate is due to the fact that
the proposed probability distribution for $z_i$ (light gray) is very
different from the accepted distribution (the Gaussian). 
}
\label{1probaaccept}
\end{figure}

To really understand what is going on in the coupled spring problem, 
let's go back to figure~\ref{1oscillators}, and 
analyze a part only of the whole game: the motion of the bead
labeled $i$ with $i-1$ and $i+1$ (for the moment) \textsc{immobilized}
at some values $z_{i-1}$ and at $z_{i+1}$. It is easy to do this simulation
(as a thought experiment) and to see that the results obtained are
as given in figure~\ref{1probaaccept}. 
You see that the distribution function  $P(z_i)$,
(at fixed $z_{i-1}$ and $z_{i+1}$) is a Gaussian centered around
$\overline{z}_i = (z_{i-1} + z_{i+1})/2$. 
Note however that there are in fact two distributions:
the accepted one and the rejected  one. It is the Metropolis algorithm which,
through the rejections, modifies the proposed - flat - distribution into the
correct Gaussian. 
We see that  the rejections - besides being a nuisance -  play the very
constructive role of modifying the proposed moves into the correct
probability density.
There is 
a whole research literature on the use of rejection methods to sample
$1-$dimensional distributions (cf \cite{Press}, chap 7.3),   
\ldots a subject which we will leave instantly
because we are more interested in the higher-dimensional case.

\section{A priori Probability}

Let us therefore  extend the 
usual formula for the detailed balance condition and for the Metropolis
algorithm, by taking into account the possible ``\textsl{a priori}'' choices of 
the moves, which is  described by an 
\textsc{a priori Probability} $\mathcal{A}(a\rightarrow b)$ to attempt the 
move \footnote{I don't know who first used this concept. I learned it 
from D. Ceperley}.
In the heliport game, this probability was simply
\begin{equation}
\mathcal{A}(a\rightarrow b) = 
  \left\{ \begin{array}{ll}
   1 & \mbox{if}\;\; |a - b| < \epsilon\\
       0 & \mbox{otherwise}
                 \end{array}
        \right.
\label{simpleapriori}
\end{equation}
with $\epsilon$ the throwing range, 
and we did not even notice its presence. In the elastic spring example, 
the probability to pick a bead $i$, and to move it by a small amount 
$-\epsilon<\delta<\epsilon$ was also independent of $i$, and of the
actual position $z_i$.

Now, we reevaluate the detailed balance equation, 
and allow explicitly for an algorithm: The probability $p(a \rightarrow b) $
is split up into two separate parts \cite{Ceperley}:
\begin{equation}
p(a\rightarrow b) = \mathcal{A}(a\rightarrow b) \mathcal{P}(a\rightarrow b)
\end{equation} 
where $\mathcal{P}(a\rightarrow b)$ is the (still necessary) acceptance 
probability of the move proposed with $\mathcal{A}(a\rightarrow b)$. 
What is this rejection probability? This is very easy to obtain from
the full detailed balance equation
\begin{equation}
\pi(a)  \mathcal{A}(a\rightarrow b)\mathcal{P}(a\rightarrow b) =
\pi(b)  \mathcal{A}(b\rightarrow a)\mathcal{P}(b\rightarrow a)
\end{equation}
You can see, that for any \textsl{a priori} probability, i.e. for \textsc{any} 
Monte Carlo algorithm we may find the acceptance rate 
which is needed to bring this
probability into accordance with our unaltered detailed balance condition.
As before, we can use a Metropolis algorithm to obtain (one possible) correct
acceptance rate
\begin{equation}
\frac{\mathcal{P}(a\rightarrow b)}{\mathcal{P}(b\rightarrow a)} =
\frac{\pi(b)}{\mathcal{A}(a\rightarrow b)}
\frac{\mathcal{A}(b\rightarrow a)}{\pi(a)}
\end{equation}
which results in 
\begin{equation}
\mathcal{P}(a\rightarrow b) = \min \left(1, 
\frac{\pi(b)}{\mathcal{A}(a\rightarrow b)}
\frac{\mathcal{A}(b\rightarrow a)}{\pi(a)} \right)
\label{Metropmod}
\end{equation}
Evidently, this formula reduces to the original Metropolis prescription
if we introduce a flat \textsl{a priori} probability 
(as in eq.~(\ref{simpleapriori})).
As it stands, eq.~(\ref{Metropmod}) 
states the basic algorithmic liberty  which we have in 
setting up our Monte Carlo algorithms: Any possible bias 
$\mathcal{A}(a\rightarrow b)$ can be made into a valid Monte Carlo algorithm 
since we can always correct it with the corresponding acceptance rate
$\mathcal{P}(a\rightarrow b)$. Of course, only a very carefully chosen
probability will be viable, or even superior to the simple and popular
choice eq.~(\ref{Metrop}). 
 
Inclusion of a general \textsl{a priori} probability is mathematically harmless, 
but generates a profound change in the practical setup of our simulation.
In order to evaluate the acceptance probability in eq. (\ref{Metropmod}), 
$\mathcal{P}(a\rightarrow b)$, we not only propose the move to $b$, 
but also need explicit evaluations of both
$\mathcal{A}(a\rightarrow b)$ \textsc{and} of the return move
$\mathcal{A}(b\rightarrow a)$. 

Can you see what has changed in the eq.~(\ref{Metropmod}) with respect to the 
previous one (eq.~(\ref{Metrop}))? Before, we necessarily had a large 
rejection probability whenever getting from a point $a$ with high
probability (large $\pi(a)$) to a point $b$ which had a 
low probability (small $\pi(b)$). The 
naive Metropolis algorithm could only produce the correct probability
density by installing rejections. 
\textsc{Now} we can counteract, by simply choosing an \textsl{a priori} probability 
which is also much smaller. In fact, you can easily see that there is an 
an optimal choice: we may be able to  use as an \textsl{a priori} probability 
$\mathcal{A}(a\rightarrow b)$ the probability density $\pi(b)$ and
$\mathcal{A}(b\rightarrow a)$ the probability density $\pi(a)$.
In that case, the ratio expressed in eq. (\ref{Metropmod}) will always 
be equal to $1$,
and there will never be a rejection. Of course,  we are then also 
back to direct sampling \ldots where we came from 
because direct sampling was too difficult \ldots .

The argument is not circular, as it may appear, because it can always be
applied to a part of the system. To understand this point, it is best to 
go back to the example of the elastic string.
We know that the probability distribution 
 $\pi(z_i)$ for fixed $z_{i-1}$ and $z_{i+1}$ is 
\begin{equation}
\pi(z_i|z_{i-1},z_{i+1}) \sim \exp[-\beta (z_i -\overline{z}_i)^2]
\label{1move}
\end{equation}
with $\overline{z}_i = (z_{i-1} + z_{i+1})/2$. We can now use 
exactly this formula as an \textsl{a priori} probability 
$\mathcal{A}( z_i|z_{i-1},z_{i+1})$ and generate an algorithm without
rejections, which thermalizes the bead $i$ at any step with its 
immediate environment. To program this rule, you need Gaussian 
random numbers (cf. \cite{Press} for the popular Box-Muller algorithm).
So far, however, the benefit of our operation is essentially 
non-existent \footnote{The algorithm with this \textsl{a priori} probability
is called ``heatbath'' algorithm. It is popular in spin models, 
but essentially identical to the Metropolis method}.
 
It is now possible to extend the formula for $z_i$ at fixed 
$z_{i-1}$ and $z_{i+1}$ to a larger window. Integrating over
$z_{i-1}$ and $z_{i+1}$ in $\pi(z_{i-2},\ldots,z_{i+2})$, we find that
\begin{equation}
\pi(z_i|z_{i-1},z_{i+1}) \sim \exp[-2 \beta (z_i -\overline{z}_i)^2]
\label{2move}
\end{equation}
where, now,  $\overline{z}_i = (z_{i-2} + z_{i+2})/2$. 
Having determined $z_i$ from a sampling of eq.~(\ref{2move}), we can 
subsequently find values for $z_{i-1}$ and $z_{i+1}$ using 
eq.~(\ref{1move}). The net result of this is that we are able to 
update $z_{i-1},z_i,z_{i+1}$ simultaneously. The program ``\texttt{levy.f}''
which implements this so called L\'{e}vy construction can be retrieved 
and studied from 
my WWW-site. It generates large moves with arbitrary window size 
without rejections.

\section{Perturbations}
From this simple example of the coupled spring problem, you 
can quickly reach all the subtleties of the Monte Carlo method. 
You see that we were able to produce a perfect algorithm, because
the \textsl{a priori} probability $\mathcal{A}( z_i|z_{i-l},z_{i+l})$  could
be chosen equal to 
the stationary probability $\pi(z_{i-l},\ldots,z_{i+l})$ resulting
in a vanishing rejection rate. This, however, was just a happy accident.
The enormous potential of the 
\textsl{a priori} probability resides in the  fact that eq.~(\ref{Metropmod})
deteriorates (usually) very gracefully when $\mathcal{A}$ and $\pi(z)$ 
differ. A recommended way to understand this point consists
in  programming a second program for the 
coupled chain problem, in which you again add a little perturbing term
to the energy, such as 
\begin{equation}
E_1 = \gamma \sum_{i=1}^{N} f(z_i)
\end{equation}
which is supposed to be relatively less important than the elastic
energy. It is interesting to see in which way the method adapts if we keep  
the above L\'{e}vy-construction as an algorithm\footnote{In Quantum Monte
Carlo, you introduce a small coupling between several strings.}.
If you go through the following argument, and possibly even write 
the program and experiment with 
its results, you will find the following
\begin{itemize}
\item The energy of each configuration now is  
$\tilde{E}(z_1,\ldots,z_N)= E_{0} + E_1$,
where $ E_{0}(a)$ is the term given in eq.~(\ref{oscillators}), which is
in some way neutralized by the \textsl{a priori} probability  
$\mathcal{A}(a \rightarrow b) = \exp[-\beta E_0(b)]$.
One can now  write down
the Metropolis acception rate of the process from eq.~(\ref{Metropmod}).
The result is
\begin{equation}
\mathcal{P}(a\rightarrow b) = \min \left(1, 
\frac{\exp[-\beta E_1(b)]}{\exp[-\beta E_1(a)]} \right)
\label{Metropmodappl}
\end{equation}
This is exactly the naive Metropolis algorithm eq.~(\ref{Metrop}), 
but exclusively with the newly added term of the energy.
\item Implementing the \textsl{a priori} probability with $\gamma=0$, 
your code runs with acceptance probability $1$, independent of the 
size of the interval $2 l + 1$. If you include the second term $E_1$, 
you will again have to optimize this size. Of course, you will 
find that the optimal 
window corresponds to a typical size of $\beta| E^b_1 - E^a_1|\sim 1$.
\item with this cutting-up of the system into a part which you can 
solve exactly, and an additional term, you will find out that the    
Monte Carlo algorithm has the appearance of a
perturbation method. Of course, it will always
be exact. It has all the chances to be fast if $E_1$ is  typically smaller
than $E_0$. One principal difference with perturbation methods is that
it will always sample the full  perturbed distribution $\pi(z_1,\ldots,z_N)=
\exp[-\beta \tilde{E}]$.  
\end{itemize}  

One can spend an essentially endless time pondering about the  
\textsl{a priori} probabilities, 
and the similarities and differences with perturbation theory.
This is where the true power of the Monte Carlo method lies. This is what
one has to understand before venturing into  custom tailering
tomorrow's powerful algorithms for the problems which are today 
thought to be out of reach. Programming a simple algorithm for the 
coupled oscillators will be an excellent introduction into the subject.
Useful further literature are \cite{Ceperley}, where the coupled
spring problem is extended into one of the most successful applications
to Quantum Monte Carlo methods, and \cite{Sokal}, where some limitations
of the approach are outlined. 

In any case, we should understand that a large rate  of 
rejections is always indicative of a structure of the proposed moves
which is unadapted to the probability density of the model at the 
given temperature. The benefit of fixing this problem - if we only
see how to do it - is usually tremendous: doing the small movements
with negligible rejection rate often allows us to do larger movements, and 
to explore the important regions of configuration space all the more
quickly. 

To end this section, I will give another example: the celebrated 
cluster methods in lattice systems, which were introduced ten years ago 
by Swendsen and Wang \cite{Swendsen} and by Wolff \cite{Wolff}. 
There again, we find the two 
essential ingredients of slow algorithms: necessarily small moves, and 
a wide distribution of the physical quantities. Using the concept of 
\textsl{a priori} probabilities, we can very easily understand these methods. 

\section{Cluster Algorithms}

\begin{figure}[htbp] \unitlength 1cm
\begin{center}
\begin{picture}(6.,5.)
\put(0.0,0){\epsfbox{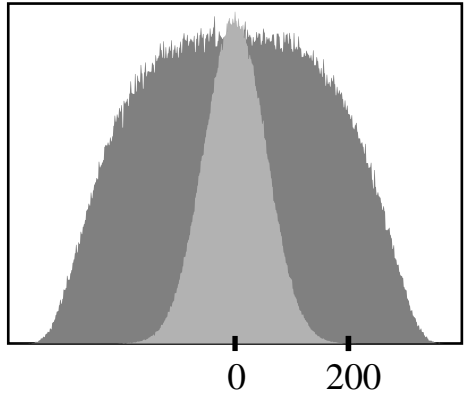}}
\end{picture}
\end{center}
\caption
{Histograms of the magnetization in the $20 \times 20$ Ising model
at temperatures 10 \% and 45 \% above the critical point. 
On the x-axis is plotted $m/\Delta m$, where $\Delta m= 2$ is the
change of the magnetization which can be obtained in a single  
Monte Carlo step. 
}
\label{1histogram}
\end{figure}

I will discuss 
these methods  in a way which 
brings out clearly the ``engineering'' aspect of the \textsl{a priori} 
probability, where one tries to cope with the large distribution problem.
Before doing this, let us discuss, as before,  the general 
setting and the physical origin of the slow convergence.
As we all know, the Ising model is defined as a system of spins on a 
lattice with an energy of
\begin{equation}
E = -\sum_{<i,j>} S_i S_j
\end{equation} 
where the spins can take on values of $S_i = \pm 1$, and where the 
sum runs over pairs of neighboring lattice sites. 
A simple program is again written in a few lines: it picks a spin 
at random, and tries to turn it over. Again, the \textsl{a priori} probability
is flat, since the spin to be flipped is chosen arbitrarily.
You may find such a program (``\texttt{simpleising.f}'')
on my WWW site. 
Using this program, you will very easily be able to 
recover the phase transition between a paramagnetic and a ferromagnetic
phase,  which in two dimensions takes place at a temperature of 
$\beta=\log(1+\sqrt{2})/2$ (you may want to look up the classic reference
\cite{Ferdinand} for exact results on finite lattices). 
You will also find out that the program is increasingly 
slow around the critical temperature. 
This is usually interpreted as the effect of 
the divergence of the correlation length 
as we approach the critical point. 
In our terms, we understand
equally well this slowing-down: our simple
algorithm changes the magnetization by at most a value of $2$ at each step, 
since the
algorithm flips only single spins.
This discrete value replaces the $\epsilon$ in our previous example of the 
coupled springs. If we now plot histograms of the total magnetization of the 
system (in units of the stepsize $\Delta m =2$!), we again see that this 
distribution becomes ``wide'' as we approach $\beta_c$ (cf 
figure~\ref{1histogram}). 
Clearly, the total 
magnetization has a wide distribution, which  it is extremely difficult
to sample with a single spin-flip algorithm. 

\begin{figure}[htbp] \unitlength 1cm
\begin{center}
\begin{picture}(11,4.5)
\put(0.0,0){\epsfbox{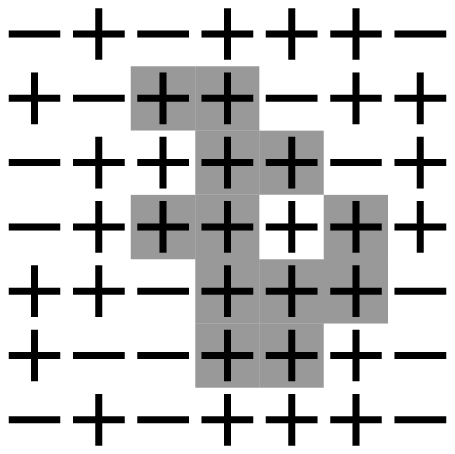}}
\put(6.5,0){\epsfbox{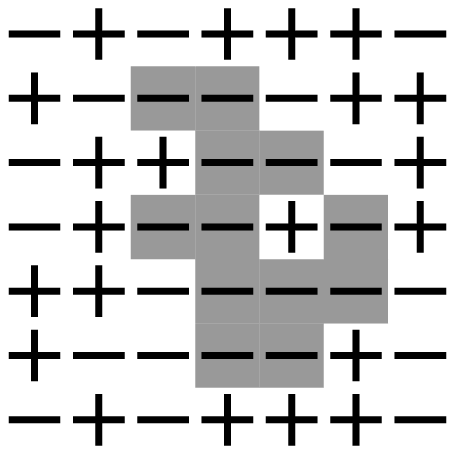}}
\end{picture}
\end{center}
\caption
{construction process of the Swendsen-Wang algorithm. Starting from an
initial $+$ site, we include other $+$ sites with probability $p$ (left). 
The whole cluster (gray area) is then flipped. In 
the text, we compute the probability to stop the construction for the 
gray cluster, and for the reverse move. This yields our \textsl{a priori} probabilities.
}
\label{1swendsen}
\end{figure}

To appreciate the cluster algorithms, you have to understand two 
things: 
\begin{enumerate}
\item  As in the coupled spring problem, you cannot simply flip several spins
simultaneously (cf the discussion on page~\pageref{simultaneous}.) 
You want to flip large
clusters, but on the other hand you cannot simply solder together all the 
spins of one sign which are connected 
to each other, because those could never again get separated.
\item  If you cannot solidly fix the spins of same sign with probability
$1$, you have to choose adjacent coaligned spins  with a certain probability
$p$. This probability $p$ is the construction parameter of our 
\textsl{a priori} probability $\mathcal{A}$. The algorithm will run for an 
arbitrary value of $p$ ($p=0$ corresponding to the single spin-flip 
algorithm), but the $p$ minimizing the rejections will be optimal.
\end{enumerate}
The cluster algorithm we find starts with the idea of choosing an 
arbitrary starting point, and adding ``like'' links with probability $p$.
We arrive here at the first  nontrivial example of the 
evaluation of an \textsl{a priori} probability.
Consider the figure~\ref{1swendsen}.
Imagine that we start from a ``+'' spin in the gray area of configuration
a) and add ``like'' spins. What is the \textsl{a priori} probability
$\mathcal{A}(a \rightarrow b)$ and the inverse probability
$\mathcal{A}(b \rightarrow a)$, and what are the Boltzmann weights
$\pi(a)$ and $\pi(b)$?

$\mathcal{A}(a \rightarrow b)$ is given by a term concerning interior
``+~+'' links, $\mathcal{A}_{int}(a \rightarrow b)$, which looks difficult,
and which we don't even try to evaluate, and a part concerning the boundary
of the cluster. This boundary is made up of two types of links, as 
summarized below
\begin{equation}
\fbox{
$
\begin{array}{ccc}
               \mbox{int}& \mbox{ext}    &  \mbox{number}  \\
                 + &    -    &   c_1  \\
                 + &    +    &   c_2  \\
\end{array}
$
}
\;\;\;
E|_{\partial C} = -c_2 + c_1
\label{atob}
\end{equation}
(in the example of figure~\ref{1swendsen}, we have $c_1=10$ and $c_2=14$). 
The \textsl{a priori} probability
is $\mathcal{A}(a \rightarrow b) = \mathcal{A}_{int}\times (1-p)^{c_2}$.
To evaluate the Boltzmann weight, we also abstain from evaluating terms
which don't change between a) and b): we clearly only need the 
boundary energy, which is given in eq.~(\ref{atob}). It follows that
$\pi(a) \sim\exp[-\beta(c_1-c_2)]$.
We now consider the inverse move. In the cluster b), the links across
the boundary are as follows:
\begin{equation}
\fbox{
$
\begin{array}{ccc}
               \mbox{int}& \mbox{ext}    &  \mbox{number}  \\
                 - &    -    &   c_1  \\
                 - &    +    &   c_2  \\
\end{array}
$
}
\;\;\;E|_{\partial C} = -c_1 + c_2
\label{btoa}
\end{equation}
The \textsl{a priori} probability to construct this cluster is again 
put together by an interior part, which is exactly the same as for 
the cluster in a), and a boundary part,  which is changed:
$\mathcal{A}(b \rightarrow a) = \mathcal{A}_{int}\times (1-p)^{c_1}$.
Similarly, we find that $\pi(a) \sim\exp[-\beta(c_2-c_1)]$.
It is now sufficient to put everything into the formula of the detailed
balance 
\begin{equation}
e^{-\beta [ c_1 - c_2]} (1-p)^{c_2} \mathcal{ P}(a\rightarrow b)
=
e^{-\beta [ c_2 - c_1]} (1-p)^{c_1} \mathcal{ P}(b\rightarrow a)
\end{equation}
which results in the acceptance probability:
\begin{equation}
\mathcal{ P}(a\rightarrow b) = \min(1,\frac
{e^{-\beta [ c_2 - c_1]}(1-p)^{c_1}}
{e^{-\beta [ c_1 - c_2]}(1-p)^{c_2}})
\label{generalswendsen}
\end{equation}
The most important point of this equation is \textsc{not} that it
can be simplified, as we will see in a minute, but that it is 
perfectly explicit, and may be evaluated without problem: once your 
cluster is constructed, you could simply evaluate $c_1$ and $c_2$ and 
compute the rejection probability from this equation.

On closer inspection of the formula eq.~(\ref{generalswendsen}), you
see that, for $1-p = \exp[ -2 \beta]$, the acceptance probability
is always $1$. This is the ``magic'' value implemented in the 
algorithms of Swendsen-Wang and of Wolff.

The resulting algorithm \cite{Wolff} is very simple, and follows 
exactly the 
description given above: you start picking a random spin, and 
add coaligned neighbors with probability $p$, the construction stops
once none of the ``like'' links across the growing cluster has been 
chosen. If you are interested, you may retrieve the program 
``\texttt{wolff.f}''
from my WWW-site. This program (which was written in less than an
hour) explains how to keep track of the cluster construction process.
It is amazing to see how it passes the Curie point
without any perceptible slowing-down. 

Once you have seen such a program churn away at the difficult problem 
of a statistical physics model close to the critical point you will 
come to understand what a great pay-off can be obtained from  an 
intelligent use of powerful Monte Carlo ideas.

\section{Concluding remarks on the equilibrium Monte Carlo}
We arrive at the end of the introduction to equilibrium 
Monte Carlo methods. I hope to have given a comprehensible introduction
to the way the method presents itself in most statistical physics 
contexts:
\begin{itemize}
\item The (equilibrium) Monte Carlo approach is an integration method
which converges slowly, but surely.
Except in a few cases, one is always forced to sample the stationary 
density (Boltzmann measure, density matrix in the quantum case) 
by a Markov chain approach. The critical 
issue there is the correlation time, which can become astronomical.
In the usual application, one is happy with a very small number of 
truly independent samples, but an appallingly large number of computations
never succeed in decorrelating the Markov chain from the initial condition.
\item The regular methods work well, but have some important disadvantages.
As presented, the condition that the rejection rate has to be quite
small - typically of the order of $1/2$ - reduces us to very local moves.
\item  The acceptance rate has important consequences for the speed
of the program, but a small acceptance rate is in particular an indicator
that the proposed moves are inadequate.
\item  It would be wrong to give the impression that the situation can 
always be ameliorated - sometimes one is simply forced to do very 
big calculations. In many cases however, a judicious choice of the 
\textsl{a priori} probability allows us to obtain very good acceptance rates, 
even for large movements.
This work is  {\em important}, and it frequently leads to an exponential 
increase in efficiency.
\end{itemize}

\chapter{Dynamical Monte Carlo Methods}
\section{Ergodicity}
Usually, together with the fundamental concept of detailed balance,
one finds also a discussion of \textsc{Ergodicity}, since it is the 
combination of both principles which insures  that the simulation will
converge to the correct probability density. Ergodicity simply means
that any configuration $b$ can eventually be reached from the initial 
configuration 
$a$, and we denote it by $p(a \rightarrow \ldots \rightarrow b) > 0$.

Ergodicity is a tricky  concept, which  does not have the step-by-step
practical meaning of the detailed balance condition. Ergodicity can be 
broken in two ways:
\begin{itemize}
\item trivially, when your Monte Carlo  dynamics for some reasons 
only samples part of phase space, split off, {\em e.g.},  by a symmetry 
principle.
An amusing example is given by the 
puzzle game we considered already in section \ref{toddler} 
\label{ergodbreak}:
It is easy to convince yourself that the 
toddlers' algorithm is \textsc{not} equivalent to the grown-up algorithm,
and that it creates only half of the possible configurations. Consider
those configurations of the puzzle with the empty square in the 
same position, say in the lower right corner, as in figure~\ref{2puzzle}.
\begin{figure}
\begin{center}
\includegraphics{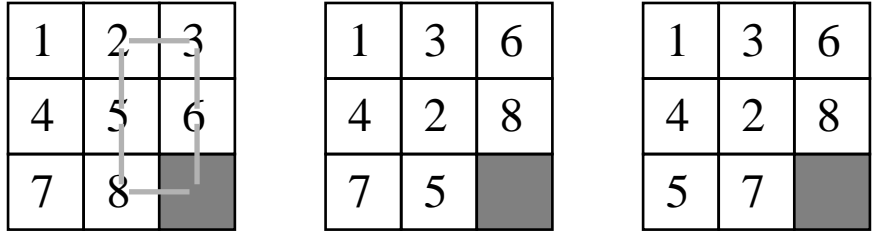}
\end{center}
\caption{The two configurations to the left  can
reach each other by the path indicated (the arrows indicate the 
(clockwise) motion of the empty square). The rightmost configuration cannot
be reached since it differs by only one transposition from the middle one.
}\label{2puzzle} 
\end{figure} 
Two such configurations 
can only be obtained from each other if the empty square describes
a closed path, and this invariably corresponds to an \textsc{even} 
number of steps 
(transpositions) of the empty square. The first two configurations
in figure~\ref{2puzzle} can be obtained by such a path. The third 
configuration (to the right) differs from the middle one in only one
transposition. It can therefore not be obtained by a local algorithm. 
Ergodicity 
breaking of the present type is very easily fixed, by simply considering
a larger class of possible moves.

\item The  much more vicious ergodicity breaking appears when the 
algorithm is ``formally correct'', but is simply too slow. After a 
long simulation time, the algorithm may not have visited all the 
relevant corners of configuration space. The algorithm may be too 
slow by a constant factor, a factor $N$, or $\exp(N)$ \ldots.
Ergodicity breaking of this
kind sometimes goes unnoticed for a while, because it may
show up clearly only in particular variables,
{\em etc}. Very often, the system can be solidly installed in some
local equilibrium, which does however not correspond to the thermodynamically
stable state. 
It  always invalidates
the Monte Carlo simulation. There are many examples of this type of ergodicity 
breaking, {\em e. g.} in the study of disordered systems. Notice that 
the abstract
proof that no ``trivial'' accident happens does not protect you from 
a ``vicious'' one. 
\end{itemize}

For \textsc{Orientation} (and knowing that I may add to the confusion)
I would want to warn the reader to think that in a benign simulation 
\textsc{all} the 
configuration have a reasonable chance to be visited. This is not at
all the case, even in small systems. Using the very fast algorithm
for the Ising model which I presented in the last chapter, you may
for yourself generate 
energy histograms of, say,  the $20 \times 20$ Ising
model at two slightly different temperatures (above the Curie point).
Even for very small differences in temperature, there are 
many configurations which have no practical chance to ever
be generated at one temperature, but which are commonly encountered at
another temperature. This is of course the legacy of the Metropolis
method where we sample configurations according to their statistical
weight $\pi(x)\sim \exp[-\beta E(x)]$. 
This socalled ``Importance Sampling'' is the rescue
(and the danger) of the Monte Carlo method - but is related only to
the equilibrium properties of the process.
Ergodicity breaking - a sudden slowdown of the simulation - 
may be completely unrelated to changes in 
the equilibrium properties of the system.  
There are a few models  in which this problem is discussed. In my 
opinion, none is as
accessible as the work of ref \cite{Sethna} which concerns 
a modified Ising model, which undergoes a purely dynamic roughening transition. 

We retain that the absolute probability  
$p(a \rightarrow \ldots \rightarrow b) > 0$
can very rarely be evaluated explicitly, and that the formal mathematical
analysis is useful only to detect the ``trivial'' kind of ergodicity 
violation. Very often, careful data analysis and much physical insight 
is needed to assure us of the practical correctness of the algorithm.

\subsection{Dynamics}
One is of course very interested in the numerical study of the phenomena
associated with very 
slow dynamics, such
as relaxation close to phase transitions, as glasses, spin glasses,
{\em etc}. We have just seen that these systems are 
{\em a priori} difficult to study with
Monte Carlo methods, since the stationary distribution is never
reached during the simulation. 

It is characteristic of the way 
things go in Physics that - nonwithstanding these difficulties - 
there is often no better method to study
these systems 
than to do a \ldots  Monte Carlo simulation! 
In \textsc{Dynamical Monte Carlo}, one is of course bound 
to a \textsc{specific} Monte Carlo algorithm, which serves as a 
\textsc{model} for the temporal evolution of the system from one 
local equilibrium state to another.
In these cases, one knows by construction that the 
Monte Carlo algorithm will drive the system towards the equilibrium, 
but very often after a time much too large to be relevant for experiment 
and simulation.
So one is interested in studying relaxation of the system from a given 
initial configuration.

The conceptual difference of the equilibrium Monte Carlo (which was
treated in the last chapter) with the  dynamical Monte Carlo methods  
cannot be overemphasized. In the first one, we have an essentially
unrestricted choice of the algorithm (expressed by the 
\textsl{a priori} probability, which was discussed at length), since one is
exclusively interested in generating independent configurations $x$
distributed according to $\pi(x)$.  In the thermodynamic Monte Carlo, 
the temporal correlations are just a nuisance. As we turn to 
dynamical calculations, these correlations become the main object of 
our curiosity. In dynamical simulations, the \textsl{a priori} probability is 
of course fixed. Also, if the Monte Carlo algorithm is ergodic both 
in principle and in practice, the static results will naturally 
be independent of the algorithm.

You may ask whether there is any direct relationship between the Metropolis
dynamics (which serves as our \textsc{Model}), and the true 
physical time of the experiment, which would
be obtained by a numerical integration of the equations of motion,
as is done for example in molecular dynamics.
There has been a lot
of discussion of this point and many simulations have been devoted
to an elucidation of this question 
for, say, the hard sphere liquid.  All these studies
have confirmed our intuition (as long as we stay with purely local 
Monte Carlo rules): the difference between the two 
approaches corresponds to a renormalization of time, as soon as 
we leave a ballistic regime (times large compared to the 
mean-free time). The Monte Carlo dynamics 
is very often simpler to study.

In equilibrium Monte Carlo, theory does not stop with the naive 
Metropolis algorithm. Likewise, in 
dynamical simulation there is also room for much algorithmic subtlety.
In fact, even though our \textsc{model} for the dynamics is fixed, we are not
forced to implement the Metropolis rejections blindly on the computer.
Again, it's the rate of rejections which serves as an important indicator
that something more interesting than the naive Metropolis algorithm
may be tried. 
The keyword here are \textsc{faster than the Clock algorithms} which 
are surprisingly little appreciated, even though they often allow simulations 
of unprecedented speed.

\section{Playing dice}
\label{dice}
\begin{figure}
\begin{center}
\includegraphics{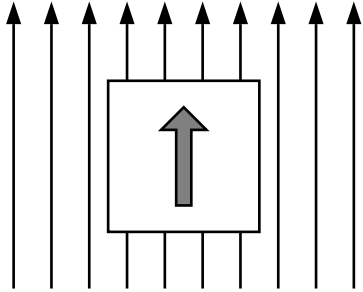}
\end{center}
\caption{Here, we look at the dynamics of a single spin in a 
magnetic field. We will soon relate it \ldots
}\label{2spin} 
\end{figure}

As a simple example which is easy to remember, consider the 
system shown in figure~\ref{2spin}: 
a single spin, which can be $S = \pm 1$ in 
a magnetic field $\mathcal{H}$, at finite temperature. The energy of 
each of the two configurations is 
\begin{equation}
               E = - \mathcal{H}\; S
\end{equation}
We consider the Metropolis algorithm of eq.~(\ref{Metrop}) to model the
dynamical evolution of the system and introduce an explicit time step
$\Delta \tau = 1$.
\begin{equation}
p(S \rightarrow -S,\Delta\tau) = \Delta\tau
\left\{ \begin{array}{cl}
              1                            & (\mbox{if}\; E(-S) < E(S)) \\
\exp[-\beta( E(-S) - E(S))] & (\mbox{otherwise})
\end{array} \right.
\label{dynamicmonte}
\end{equation}
To be completely explicit, we write down again how the spin state for
the next time step is evaluated in the Metropolis rejection procedure: 
at each time step $t=1,2, \ldots$
we compute a random number  $0 < ran < 1$ and compare it with the 
probability $p$ from eq. (\ref{dynamicmonte}):
\begin{equation}
      S_{t+1}=\left \{ \begin{array}{cl}
-S_t & \mbox{if}\; p(S_t \rightarrow -S_t) > \mbox{ran} \\
 S_t & \mbox{otherwise}
\end{array}
\right.
\label{dynreject}
\end{equation}
This rule assures that, asymptotically, the 
two configurations are chosen with the Boltzmann distribution.
Notice that, whenever  we are in the excited ``$-$'' state of the 
single spin model, our probability to fall back on the next 
time step is $1$, which means that the system will flip back to the 
``$+$'' state on the next move. Therefore, the simulation will look 
as follows:
\begin{equation}
\ldots + + + + \framebox{- +} +  \framebox{- +} + +  \framebox{- +}  
+ + + + \framebox{- +}
 + + \framebox{- +}  \ldots
\end{equation}
As the temperature is lowered, the probability to be in the 
excited state will gradually diminish, and you will spend more and
more time computing random numbers in eq.~(\ref{dynreject}), 
but rejecting the move  from $S=+1$ to $S=-1$.

\begin{figure}[htbp] \unitlength 1cm
\begin{center}
\begin{picture}(6.8,6.)
\put(0.0,1.){\epsfbox{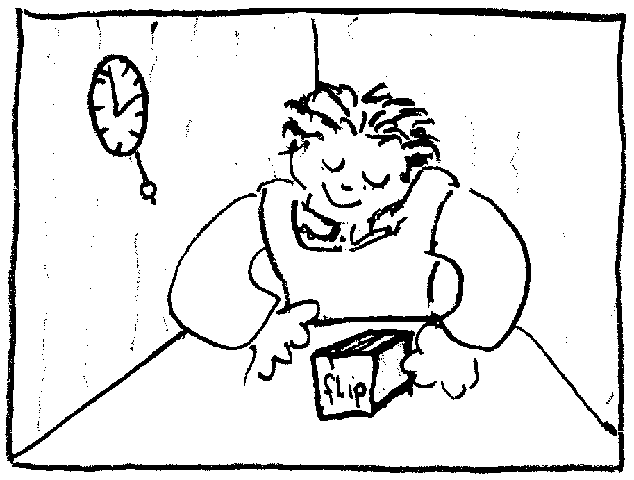}}
\put(0.2,0.){\epsfbox{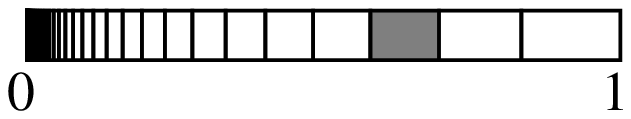}}
\end{picture}
\end{center}
\caption{\ldots to the game of this little boy. There is a simpler way
to simulate the dynamics of a single spin than to throw a die at
each time step. The solution is indicated in the lower part: the shaded
space corresponds to the probability of obtaining a ``flip'' on the 
third throw.}\label{2boy} 
\end{figure}

At the temperature $\beta=\log(6)/2$, the probability to flip the spin 
is exactly $1/6$, and the physical problem is then identical to 
the game of the little boy depicted in figure \ref{2boy}. 
He is playing with
a special die, with 
$5$ empty faces (corresponding to the rejections $S_{t+1} = S_t=1$) 
and one face with the inscription ``flip'' ($S_{t+1} =- S_t$, 
$S_{t+2} =- S_{t+1}$). 
The boy will roll the die over and over again, but of course most of the 
time the result of the game is negative. 
As the games goes on, he will
expend energy, get tired, etc etc, mostly
for nothing. Only very rarely does he encounter the magic face which 
says ``flip''. If you play this game in real life or on the computer, 
you will soon get 
utterly tired of all these calculations which result in a rejection,
and you may get the idea that there must be a more economical way to 
arrive at the same result.
In fact, you can predict analytically what 
will be the distribution of  time intervals between  ``flips''. For the 
little boy, at any given time, there is a probability of $5/6$ that one
roll of the die will result in a rejection, and a probability of 
$(5/6)^2$ that two rolls result in successive rejections, etc. The numbers
$(5/6)^l$ are shown in the lower part of figure~\ref{2boy}. You can 
easily convince yourself that the shaded space in the figure corresponds
to the probability $(5/6)^2 - (5/6)^3$ to have a flip at exactly the 
third roll.
So, to see how many times you have to wait until obtaining a ``flip'',
you simply draw a random number ran $0 < ran<1$, and check into which 
box it falls. The box index $l$ is calculated easily:
\begin{equation}
(5/6)^{l+1} < ran < (5/6)^{l}  \Rightarrow l = 
\mbox{Int}\left\{\frac{\log{ran}}{log(5/6)} \right\}.
\label{boxforl}
\end{equation}
What this means is that there is a 
very simple program ``\texttt{dice.f}'', which you can obtain from my WWW-site, 
and which has the following characteristics:
\begin{itemize}
\item the program has no rejection. For each random number drawn, the program
computes a true event: the waiting time for the next flip.
\item The program thus is ``faster than the clock'' in that it is able
to  predict the state of the boy's log-table at simulation time t
with roughly t/6 operations.
\item the output of the accelerated program is completely indistinguishable
from the output of the naive Metropolis algorithm.  
\end{itemize}
You see that to determine the temporal evolution of the die-playing 
game you don't  have to do a proper ``simulation'', you can use
a far better algorithm.

It is of course possible to generalize the little example from a 
probability $\lambda= 1/6$ to a general value of $\lambda$, and
from a single spin to a general statistical mechanics problem with 
discrete variables. \ldots I hope you will remember the trick next time
that you are confronted to a simulation, and the terminal output
indicates one rejection after another. So you will remember that 
the proper rejection method eq.~(\ref{dynreject}) is just  one 
possible implementation 
of the Metropolis algorithm eq.~(\ref{dynamicmonte}).
The method has been used to study 
different versions of the Ising model, disordered spins, 
kinetic growth and many other phenomena.
At low temperatures, when the rejection probabilities increase,
the benefits of this method can be enormous.

So, you will ask why you have never heard of this trick before. One of the 
reason for the relative obscurity of the method can already be seen 
on the one-spin example:  In fact you see that the whole method 
does not actually use the factor $1/6$, which is the probability to 
do something, but with $5/6 = 1- 1/6$, which is the probability to do 
nothing.  
In a general spin model, you can flip each of the $N$ spins. 
As input to your algorithm computing waiting times, you again need
the probability ``to do nothing'', which is 
\begin{equation}
1 - \lambda = 1 - \sum_{i=1}^N [\mbox{probability to flip spin $i$}]
\end{equation} 
If these probabilities all have to be computed anew for each new motion,
the move becomes quite expensive (of order $N$). 
A straightforward implementation 
therefore has all the chances to be too \textsc{onerous} to be extremely
useful. In practice, however, you may encounter enormous simplifications
in evaluating $\lambda$ for two reasons:
\begin{itemize}
\item you may be able to use symmetries to compute all the probabilities.
Since the possible probabilities to flip a spin fall into a finite
number $n$ of \textsc{classes}. The first paper on accelerated 
dynamical Monte Carlo algorithms \cite{Bortz} has coined the name 
``n-fold way'' for this strategy.
\item You may be able to simply \textsc{look up} the probabilities, instead
of computing them, since
you didn't visit the present configuration for the first time
and you took notes \cite{Mezard}.
\end{itemize}

\section{Accelerated Algorithms for Discrete Systems}

We now  
discuss the main characteristics of the method, as applied to a general 
spin model with configurations $S_i$ made up of $N$ spins 
$S_i= (S_{i,1},\ldots,S_{i,N})$. We denote the configuration obtained
from $S_i$ by flipping the spin $m$ as $S_i^{[m]}$. The system
remains in configuration $S_i$ during a time $\tau_i$,
so that the time evolution can be written as 
\begin{equation}
S_1(\tau_1) \rightarrow S_2(\tau_2)
 \ldots \rightarrow S_i(\tau_i)
 \rightarrow S_{i+1}(\tau_{i+1})\ldots
etc
\label{MCsequence}
\end{equation}
The probability of flipping spin $m$ is given by  
\begin{equation}
p(S_i \rightarrow S_i^{[m]},\Delta\tau) = \frac{\Delta\tau}{N}
\left\{ \begin{array}{cl}
1                        & (\mbox{if}\;\; E(S_i^{[m]}) < E(S_i)) \\
\exp[-\beta( \Delta E)]  & (\mbox{otherwise})
\end{array} \right. 
\label{transition}
\end{equation}
where the $1/N$ factor expresses the the $1/N$ probability to select the 
spin $m$.
After computing $\lambda = \sum_i p(S_i \rightarrow S_i^{[m]})$, we 
obtain the waiting time as in eq. (\ref{boxforl}), which gives the 
exact result for a finite values of $\Delta \tau$. Of course, in the 
limit $\Delta \tau \rightarrow 0$, the eq.~(\ref{boxforl}) simplifies,
and we can sample the time to stay
in $S_i$ directly from an exponential distribution
$p(\tau_i)=\lambda \exp(-\lambda \tau_i)$ (cf \cite{Press} chap. 7.2 
for how to 
generate exponentially distributed random numbers).

If we have then found out that after a time $\tau$ we are going to move
from our configuration $S$, \textsc{where} are we going? The answer 
to this question can be simply understood by looking at  
figure~\ref{2pile}: 
\begin{figure}[htbp] \unitlength 1cm
\begin{center}
\begin{picture}(5.1,4.9)
\put(0.0,0){\epsfbox{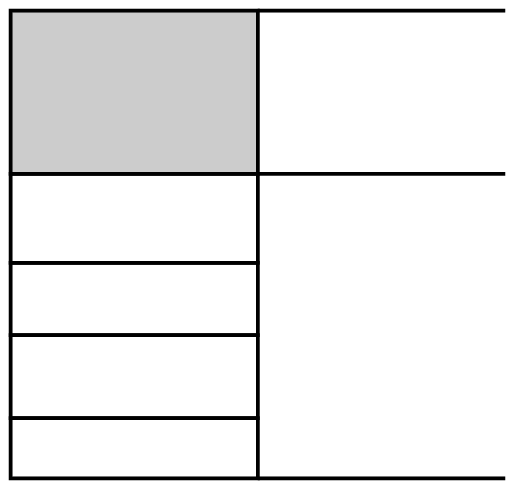}}
\put(.3,.2) {\mbox{$p(S\rightarrow S^{[1]})$}}
\put(.3,.85){\mbox{$p(S\rightarrow S^{[2]})$}}
\put(.3,2.4){\mbox{$p(S\rightarrow S^{[N]})$}}
\put(.5,3.7){\mbox{{\large $1-\lambda$}}}
\put(2.8,.7){\mbox{$\leftarrow ran$}}
\end{picture}
\end{center}
\caption
{``Pile'' of probabilities, which allows us to understand how we decide
on the spin to flip. Accelerated algorithms make sense if, usually, 
$\lambda <<1$. }
\label{2pile}
\end{figure}
there we see the ``pile'' of all the probabilities which were computed.
If the waiting time $\tau_i$ is obtained from $\lambda_i$, we choose the 
index $[m]$ of the flipped spin with a probability 
$p(S \rightarrow S^{[m]})$. 
To do this, 
you need all the elements to produce the box in 
figure~\ref{2pile} i. e. the probabilities:
\begin{equation}
\begin{array}{l}
p(S \rightarrow S^{[1]})+p(S \rightarrow S^{[2]})+\ldots
p(S \rightarrow S^{[N]})=\lambda\\
\vdots\\
p(S \rightarrow S^{[1]})+p(S \rightarrow S^{[2]})\\
p(S \rightarrow S^{[1]})\\
\end{array}
\label{pileofprob}
\end{equation}
In addition, one needs a second random number
($0<ran<\lambda$) to choose one of the boxes (cf the problem in figure
\ref{2boy}).
The best general algorithm to actually compute $m$ is of course not 
``visual inspection of a postscript 
figure'', but what is called ``search of an ordered table''. This you find
explained in any book on basic algorithms (cf, for example \cite{Press},
chap. 3.2). Locating the correct box only takes of the order of $\log(N)$ 
operations.
The drawback of the computation is therefore that any move
costs an order of $N$ operations, since in a sense we have 
to go through all the possibilities of doing something before knowing
our probability ``to do nothing''.
This  expense is the price to be payed in order to 
completely eliminate the rejections. 

\section{Implementing Accelerated Algorithms}
Once you have understood the basic idea of accelerated algorithms, you may 
wonder how these methods are implemented, and whether it's worth the 
trouble.  
In cases in which the local energy  is the same for any lattice site, 
you will find out that the probabilities can take on only $n$ different
values. In the case of the $2$-dimensional Ising model, there are
$10$ classes that the spin can belong to, ranging from 
up-spin with $4$ neighboring up-spins, up-spin with $3$ neighboring up-spins,
to down-spin with $4$ neighboring down-spins.
Knowing the repartition into different classes at the initial time, 
you see that flipping the spin changes the classes of $5$ spins, and
can be seen as a change of the number of particles belonging to 
the different classes. Using a somewhat more intricate bookkeeping,
we can therefore compute the value of   $\lambda$ in a constant number
of operations, and the cost of making a move is reduced to order $O(1)$.
You see that the accelerated algorithm has truly solved the problem of
small acceptance probabilities which haunts so many simulations at
low temperature (for practical details, see refs~\cite{Bortz},
\cite{Novotny1}).

\begin{figure}[htbp] \unitlength 1cm
\begin{center}
\begin{picture}(4.6,4.2)
\put(0.0,0){\epsfbox{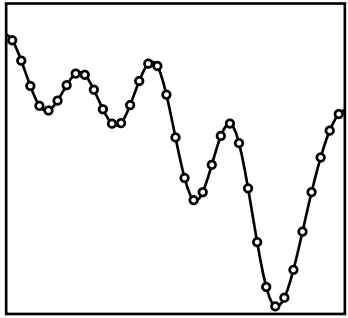}}
\end{picture}
\end{center}
\caption
{Rugged energy ``landscape'' which poses problems in a dynamical 
Monte Carlo simulation at low temperatures. The system will be
stuck in one of the local minima, and the dynamics will be futile,
i. e. very repetitive.
}
\label{2landscape}
\end{figure}

\ldots If you program the method, the impression of bliss may well turn into
frustration, for we have overlooked an important point: the system's 
dynamics, while without rejections, may still be futile.
Consider for concreteness an energy landscape as in figure~\ref{2landscape},
where any configuration (one of the little dots) is connected to two 
other configurations.
Imagine the system at one of the local minima at some initial time. 
At the next 
time step, it will move to one of the neighboring sites, 
but it will almost certainly fall back right afterwards. 
At low temperature, the system
will take a very long time (and, more importantly, a very large number
of operations) before hopping 
over one of the potential barriers. In these cases, the dynamics is 
extremely repetitive, futile. 
If such behavior is foreseeable, it is of course wasteful to recompute the 
```pile of probabilities'' eq.~(\ref{pileofprob}), and even to embark 
on the involved book-keeping tricks of the n-fold way algorithm. In these
cases, it is much more economical to
save much of the 
information about the probabilities, and to look up all the relevant 
information. An archive can be set up in such a way that, upon making the 
move $S_i \rightarrow S_i^{[m]}$ we can quickly decide whether we have
seen the new configuration before, and we can immediately look up the 
``pile of probabilities''. This leads to extremely fast algorithms
(for practical details, see \cite{Pluchery}, \cite{Mezard}).
Systems in which this  approach  can be used contain: 
flux lines in a disordered
superconductor, the NNN Ising model \cite{Sethna} alluded to earlier, 
disordered
systems with a so-called single-step replica symmetry breaking transition, 
and in general systems with steep local minima. For these systems it
is even possible to produce a second-generation algorithm, which 
not only accepts a move at every timestep, but even a move to a 
configuration which the system has never seen before. One such algorithm
has been proposed in \cite{Pluchery}, cf also \cite{Novotny2}.

There are endless possibilities to improve the dynamical algorithms
for some systems. Of course, the more  powerful a method the less 
frequently it can be applied. It should be remembered that the above 
method is only of use if the probability to do nothing is very 
large, and/or if the dynamics is very repetitive (steep local minimum
of energy function in figure~\ref{2landscape}). A large class of 
systems for which none of these conditions hold are the spin glasses
with a socalled continuous replica symmetry breaking transition, as 
the Sherrington-Kirkpatrick model. 
In these cases, the ergodicity breaking takes place very ``gracefully'', 
there are very many configurations accessible for any given initial 
condition. In this case, there seems to be very little room for
improvements of the basic method. 
 
\section{Random Sequential  Adsorption}
\label{depos}
I do not want to leave the reader  with the impression that the 
accelerated algorithms are restricted only to  spin models. In fact, intricate
rapid methods can be conceived in most cases in which you have many 
rejections in a dynamical simulation. These rejections simply 
indicate that the time to ``do something'' may be much larger than the 
simulation time step. 

\begin{figure}
\begin{center}
\includegraphics{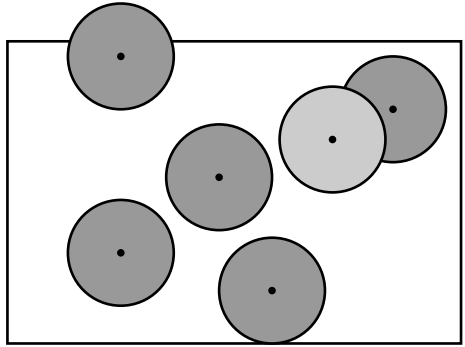}
\end{center}
\caption{Random sequential adsorption: the light coin - which has just
been deposited - has to be taken away again.
}\label{2deposit} 
\end{figure}

The following example,   
random sequential adsorption, was already mentioned in section \ref{Monaco}.
Imagine a large two-dimensional square  on 
which you deposit one  coin per second - but attention: we only put
the coin if it does not overlap with any of the coins already deposited.
The light-gray coin in the figure~\ref{2deposit} 
will immediately be taken away. 
Random sequential adsorption is a particularly
simple dynamical system because of its irreversibility.
We are interested in two questions: 
\begin{itemize}
\item The game will stop at the moment at which it is impossible 
to deposit a new coin. What is the time after which this ``jamming''
state is reached and what are its properties?
\item We would also like to know the mean density as a function of time for
a large number of realizations.
\end{itemize}
I will give here a rather complete discussion of this problem in six 
steps, displaying the panoply of refinements which can be unfolded
as we come to understand the program. There is a large research literature
on the subject, but you will see that you can yourself find an optimal
algorithm by simply applying what we have learned in the problem of the 
rolling die.
\subsection{Naive Algorithm}

You can write a program simulating in a few minutes. You need a table
which contains the ($x, y$) positions of the $N(t)$ coins already deposited, 
and a random number generator which will give the values of $x,y$.
If you run the
program, you will see that even for modest sizes of the box, it will 
take quite a long time before it stops. 
You may say ``Well, the program is slow, simply because the physical
process of random absorption is slow, there is nothing I can do \ldots''.
If that is your reaction, you may gain something from reading on. 
You will find out that there is a whole cascade of improvement that can 
be imported into the program. These improvements concern not only 
the implementation details but also the deposition process itself, which 
can be simulated faster than in the experiment - especially in the 
final stages of the deposition. Again, there is a 
\textsc{Faster than the Clock} algorithm, which deposits (almost) one
particle per unit time. These methods have been very little explored
in the past.

\subsection{Underlying Lattice}
\begin{figure}
\begin{center}
\includegraphics{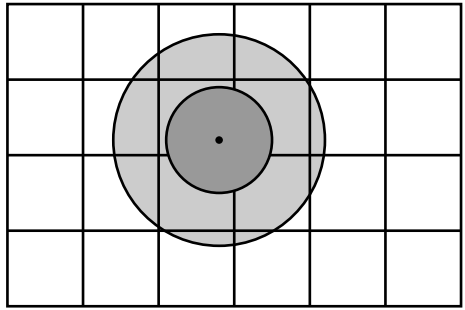}
\end{center}
\caption{One of the basic techniques to orient oneself is introducing a 
grid. The exclusion area of the coin is also shown.
}\label{2grid} 
\end{figure}

The first thing you will notice is that the program spends a lot of 
time computing distances between the proposed point ($x,y$) 
and the coins which are already deposited.
It is evident that you will gain much time by performing only a local 
search. This is done with a grid, as shown in figure~\ref{2grid}
and by computing the table of all the particles contained in each of the 
little squares. As you try to deposit the particle  at $(x,y)$, you first
compute the square which houses the point ($x,y$) and then compute the 
overlaps with particles contained in neighboring squares.
Some bookkeeping is necessarily involved, and varies with the size
of the squares adopted. 
 There has been a
lot of discussion about how big the little squares have to be taken, 
and there  is no clear-cut answer. Some people prefer a box of size 
approximately
$\sqrt{2}$ times the radius of the spheres. In this case you are sure
that at most $1$ sphere per square is present, but the number of boxes
which need to be scrutinized is quite large.
Others have adopted larger boxes which have the advantage that only
the contents of $9$ boxes have to be checked.
In any case, one gains an important factor $\mbox{N}$ with respect to 
the naive implementation.

\subsection{Stopping criterion}
Since we said that we want to play the game up to the bitter end, 
you may want to find out whether there is at all a possibility to deposit
one more coin. The best thing to do is to write a program which will 
tell you whether the square can host one more point. To do this, you have
to apply basic trigonometry to find out whether the whole square is 
covered with ``exclusion disks'', as shown in figure \ref{2grid}

\subsection{Excluding Squares}
\begin{figure}
\begin{center}
\includegraphics{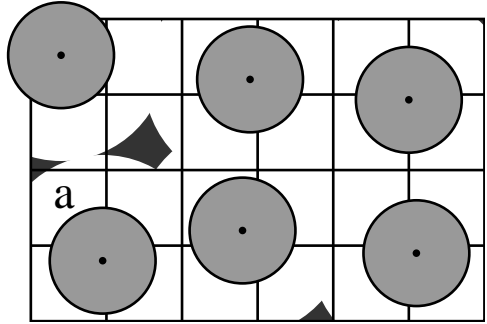}
\end{center}
\caption{Even though there is much empty space, we can only deposit 
further coins on three ``stars'' (belonging to $4$ of the $24$ squares
making up the field).
}\label{2exclude} 
\end{figure}

Try to apply the idea of an exclusion disk to the configuration shown 
in figure \ref{2exclude}.
Using the idea of the exclusion disk, you will be able to compute 
the parts of the field on which you can still deposit a new coin.
These parts have been designed in dark, we call them
``stars'' for obvious reasons. You can see that, at the present moment,
there are only $4$ of the $24$ grid-squares which can take on a new point.
Before starting to compute distances with coins in adjacent squares, it may 
be a good idea to check at all whether it is possible to deposit a box
on the point. You will quickly realize that we are confronted with 
exactly the same problem as the boy in figure~\ref{2boy}: only one out of
six grid-squares has a non-zero chance to accept a new coin.
The time for the next hit of one of the useful squares can be obtained
with a one-sides die, or, equivalently,  
sampled from eq.~(\ref{boxforl}).
So, you can write a faster (intermediate) program, by determining
which of the boxes can still hold a coin. This probability then gives 
the probability ``to do nothing'', which is used in eq.~(\ref{boxforl}) 
to sample the 
time after which a new deposition is attempted.
Are you tempted to write such a program? You simply need a 
subroutine able to determine whether there is still 
free space in a square.
With such a subroutine you are able to 
exclude 
squares from the consideration. The ratio of excluded squares to the
total number of squares then gives the probability $1-\lambda$  to 
do nothing, which is what you need for the 
faster-than-the-clock algorithm of section \ref{dice}.

\subsection{The Ultimate Algorithm}
Cutting up the field into little squares  allows us a second time 
to make the program run faster
by a factor $N$, where $N$ is the number of little squares. Not only can
we use the squares to simplify the calculation of overlaps, but to 
exclude large portions of the field from the search.

Unfortunately,
you will quickly find out that the program still has a very large rejection
probability \ldots just look at the square denoted by an \textbf{a} in 
figure~\ref{2boy}: roughly $2\%$ of the square's surface can only accept 
a new coin. So, you will attempt many depositions in vain before being 
able to do something reasonable.  One idea to go farther consists
in going to smaller and smaller squares. This has been implemented
in the literature \cite{Wang}.
What one really wants to do, however, is to exploit the exact 
analogy between the area of the stars, and the probabilities in 
eq.~(\ref{pileofprob}). If we know the location and the area of the 
stars, we are able to implement one of the rejection-free algorithms.
Computing the area of a ``star'' is a simple trigonometric exercise.
Having such a subroutine at our disposal allows us to envision the 
\textsc{ultimate program} for random sequential adsorption.
\begin{itemize}
\item Initially, you do the naive algorithm for a while
\item Then you do a first cut-up into stars. 
The total relative area of the field not covered by stars corresponds to 
the factor $\lambda$, and you will sample the 
the star for the next  deposition exactly as in 
eq.~(\ref{pileofprob}). 
Then you randomly deposit ($x,y$) into the star chosen, and update
the book keeping.
\end{itemize}

\subsection{How to sample a Random Point in a Star}

\begin{figure}
\begin{center}
\includegraphics{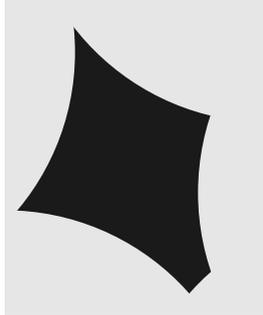}
\end{center}
\caption{ The ultimate algorithm for random sequential adsorption
needs a program to calculate the area of a star (straightforward),
and a method to sample a random point within it. Can you sample
random points within the star without rejections?
}\label{2star} 
\end{figure}

So, finally, this breath-taking, practical discussion brings us 
to ponder a philosophical question: \textsc{how to sample loci in a star}.
In fact, how do we do that?\footnote{There is no help in turning to the 
literature. the Question is neither 
treated in  ``Le Petit Prince'' by A. de St.~Exupery, nor in any other 
book I know.}. For my part, 
I have given up looking for a rejection-free method to solve this
problem - I simply sample a larger square, as in figure \ref{2star}, and
then use the good old rejection method. But perhaps you know how to do this?

\subsection{Literature, extensions}

Of course, the example of the random adsorption was only given to 
stimulate you to think about better algorithms for \textsc{Your} current
Monte Carlo problem, and how it may be possible in your own research problem 
to get away from a blind use of algorithms.  If you want
to know more about random deposition, notice that there is a vast
research literature, and an algorithm has been presented in \cite{Wang}.
Notice that in our algorithm it was very important for the 
spheres to be monodisperse, ie for them all to have the same diameter.
What can be done in the contrary case? Are there accelerated algorithms
for spheres with some distribution of diameters (from $d_{min}$ to 
$d_{max}$) (easy), and what would be an optimal algorithm for 
deposition of objects with additional degrees of freedom?  The problem
is of some interest in the case of ellipses. Evidently, from a numerical
point of view, you will end up with a three-dimensional ``star'' in 
which you have to sample ($x,y,\theta$), where $\theta$ gives the 
orientation of the ellipse  to be deposited. You may be inspired to
think about such a simulation. Remember that it is not important to 
compute the $3-d$ star exactly, just as, in the last chapter, it was
without real importance that the \textsl{a priori} probability $\mathcal{A}(x)$ 
could be made exactly to $\pi(x)$.


\begin{thebibliography}{99}
\bibitem{Metrop}
N. Metropolis, A. W. Rosenbluth, M. N. Rosenbluth, A. H. Teller, E. Teller,
{\sl J. chem. Phys} {\bf 21} 1087 (1953)

\bibitem{Binder} {\sl Monte Carlo Methods
in Statistical Physics}, edited by K. Binder, 2nd ed.
(Springer Verlag, Berlin, 1986)

\bibitem{Sokal}
S. Caracciolo, A. Pelissetto, A. D. Sokal, {\sl Phys. Rev. Lett} {\bf 72}
179 (1994)

\bibitem{Swendsen}
R. H. Swendsen and J.-S. Wang {\sl Phys. Rev. Lett.}{ \bf 63}, 86 (1987)

\bibitem{Wolff}
U. Wolff {\sl Phys. Rev. Lett.} {\bf 62}, 361 (1989)

\bibitem{Bortz} A. B. Bortz, M. H. Kalos, J. L. Lebowitz;
{\sl J. Comput. Phys.}
{\bf 17}, 10 (1975) {\it cf} also : K. Binder in \cite{Binder} sect 1.3.1

\bibitem{Sethna} J. D. Shore, M. Holzer, J. P. Sethna; {\sl Phys Rev.  B}
{\bf  46} 11376 (1992)

\bibitem{Press}
      W. H. Press, S. A. Teukolsky, W. T. Vetterling, B. P.
      Flannery, {\sl Numerical Recipes}, 2nd edition,
      Cambridge University Press (1992).

\bibitem{Ceperley}E. L. Pollock, D. M. Ceperley  {\sl Phys. Rev. B}
      {\bf 30}, 2555 (1984),  {\bf 36} 8343 (1987);
      D. M. Ceperley {\sl Rev. Mod. Phys} {\bf 67}, 1601 (1995)

\bibitem{Wang} J-S Wang {\sl Int. J. Mod. Phys C} {\bf 5}, 707 (1994)

\bibitem{Pluchery}  W. Krauth, O. Pluchery
      {\sl J. Phys. A: Math Gen} {\bf 27}, L715 (1994)

\bibitem{Mezard} W. Krauth, M. M\'{e}zard
      {\sl Z. Phys. B} {\bf 97} 127 (1995)

\bibitem{Ferdinand} A. E. Ferdinand and M. E. Fisher
      {\sl Phys. Rev.} {\bf 185} 185 (1969)

\bibitem{Lee} J. Lee, K. J. Strandburg
{\sl Phys Rev.  B}
{\bf  46} 11190 (1992)

\bibitem{Novotny1} M. A. Novotny {\sl Computers in Physics} {\bf 9} 46 (1995)
\bibitem{Novotny2} M. A. Novotny {\sl Phys. Rev. Lett.} {\bf 74} 1 (1995)
       Erratum: {\bf 75} 1424 (1995)      
\end{thebibliography}
\end{document}